\documentclass[epj]{svjour}
\usepackage{helvet}         
\usepackage{courier}        
\usepackage{type1cm}        
\usepackage{graphicx}        
\usepackage{amsmath}        %
\usepackage[bottom]{footmisc}
\usepackage{makeidx}         
\makeindex             
\begin{document}
%
\title{Relativistic Thermodynamics:\\
Transverse Momentum Distributions in High-Energy Physics.}
\author{J. Cleymans \and D. Worku}
\institute{UCT-CERN Research Centre and Department of Physics, University
of Cape Town, South Africa 
}
\date{\today}
\abstract{
Transverse momentum distributions
measured by the STAR~\cite{Abelev:2006cs} and 
PHENIX~\cite{Adare} collaborations at 
the Relativistic Heavy Ion Collider 
and by the ALICE~\cite{Aamodt:2011zj}, ATLAS~\cite{aad}
 and CMS~\cite{Khachatryan:2011tm} 
collaborations at the Large Hadron Collider can be considered
in the framework of relativistic thermodynamics 
using the Tsallis distribution.
Theoretical issues are clarified concerning the thermodynamic 
consistency in the case of 
relativistic high energy quantum distributions.    
An improved  form is proposed for describing the transverse 
momentum distribution and fits are presented together with estimates of
the parameter $q$ and the temperature $T$.
}
\PACS{25.75.Dw \and 13.85.Ni}
%
\maketitle
\section{\label{sec:Introduction}Introduction}
The Relativistic Heavy Ion Collider (RHIC) and the 
Large Hadron Collider (LHC) have produced a wealth of new data 
over the past few years. This has opened up a new energy region where
relativistic thermodynamics and hydrodynamics can be tested and applied.
The highest available energy  for heavy ions is
$\sqrt{s} = 2760$ AGeV yet the  observed temperature is only of the
order of $T \approx $ 0.16 GeV at RHIC as well as at the LHC.  
This enormous change from the energy available in the initial state to the temperature observed in the 
final state is clearly a challenge for dynamical models.

In the analysis of the new data, one statistical distribution 
has gained prominence with very good fits to the transverse momentum distributions  made
by the STAR~\cite{Abelev:2006cs} and PHENIX~\cite{Adare} collaborations at RHIC 
and by the ALICE~\cite{Aamodt:2011zj}, ATLAS~\cite{aad} and CMS~\cite{Khachatryan:2011tm} collaborations at the LHC.

In the literature there exists more than one version of the Tsallis 
distribution~\cite{Tsallis:1987eu,Tsallis:1998ab,biro_book,biro,wilk}
and we investigate here one that we consider well suited for
describing results in high energy physics. Our main guiding criterium will be thermodynamic consistency which
has not always been implemented correctly (see e.g.~\cite{Pereira:2007hp,Pereira:2009ja,Conroy:2010wt}).
The explicit form which we will use  is~\cite{c-w}: 
\begin{eqnarray}
&&\frac{d^{2}N}{dp_T~dy} =  \nonumber\\
&&gV\frac{p_Tm_T\cosh y}{(2\pi)^2}
\left[1+(q-1)\frac{m_T\cosh y -\mu}{T}\right]^{-q/(q-1)},
\label{eq:Tsallis-B}
\end{eqnarray}
where $p_T$ and $m_T$ are the transverse momentum and mass respectively, $y$
is the rapidity, $T$ and $\mu$ are the temperature and the chemical potential,
$V$ is the volume, $g$ is the degeneracy factor. In the limit where the parameter $q$ goes to 1 this reduces 
the standard Boltzmann distribution:
\begin{eqnarray}
&&\lim_{q\rightarrow 1}\frac{d^{2}N}{dp_T~dy} = \nonumber\\
&&gV\frac{p_Tm_T\cosh y}{(2\pi)^2}
\exp\left(-\frac{m_T\cosh y -\mu}{T}\right).
\label{boltzmann}
\end{eqnarray}
The parameterization given in Eq.~\eqref{eq:Tsallis-B} is close to
the one used by the STAR, PHENIX, ALICE, ATLAS and CMS
collaborations~\cite{Abelev:2006cs,Adare,Aamodt:2011zj,aad,Khachatryan:2011tm}: 
\begin{eqnarray}
  \frac{d^2N}{dp_T\,dy} = &&p_T \frac{dN}
  {dy} \frac{(n-1)(n-2)}{nC(nC + m_{0} (n-2))}\nonumber\\
&& \left[ 1 + \frac{m_T - m_{0}}{nC} \right]^{-n}
\label{ALICE-CMS}
\end{eqnarray}
where $n$, $C$ and $m_0$ are fit parameters. The analytic expression used in 
Refs.~\cite{Abelev:2006cs,Adare,Aamodt:2011zj,aad,Khachatryan:2011tm} corresponds to identifying 
\begin{equation}
n\rightarrow \frac{q}{q-1}
\end{equation}
and 
\begin{equation}
nC  \rightarrow \frac{T+m_0(q-1)}{q-1}  .
\end{equation}
After  this substitution Eq.~\eqref{ALICE-CMS} becomes
\begin{eqnarray}
  \frac{d^2N}{dp_T\,dy} =&& p_{T} \frac{{\rm d}N}
  {{\rm d}y} \frac{(n-1)(n-2)}{nC(nC + m_{0} (n-2))}\nonumber\\ 
&&\left[\frac{T}{T+m_0(q-1)}\right]^{-q/(q-1)}\nonumber\\
&&\left[ 1 + (q-1)\frac{m_T}{T} \right]^{-q/(q-1)}  .
\label{sub}
\end{eqnarray}
Which, at mid-rapidity $y=0$ and zero chemical potential,
 has the same dependence on the 
transverse momentum as~\eqref{eq:Tsallis-B} apart from an
additional  factor $m_T$. 
The inclusions of the factor $m_T$ 
leads to a more consistent interpretation of the variables $q$ and $T$.
In particular, no clear pattern emerges for the values of $n$ and $C$ while 
an interesting regularity 
is obtained for $q$ and $T$ as seen in 
Figs.~\ref{fig:q_value} and~\ref{fig:T_value} shown towards the 
end of this paper.
\\
The Tsallis distribution introduces a new parameter $q$ which in practice is always close to 1,
typical values for the parameter $q$  obtained from fits to the transverse momentum distribution 
are in the range 1.1 to 1.2, in the remainder of this paper we will always assume $q>1$.

In  section 2 we review the derivation of the Tsallis distribution by emphasizing the quantum statistical form. 
In section 3 we prove thermodynamic consistency. In section 4 we show in detail fits to the transverse momentum 
distribution in $p-p$ collisions at $\sqrt{s}$ = 900 GeV.
Section 5 presents conclusions.
%
%
%
\section{\label{sec:High}Tsallis Distribution.}
\subsection{Comparison with Standard Statistical Distributions} 
\begin{figure}
\begin{center}
\includegraphics[width=0.45\textwidth,height=6cm]{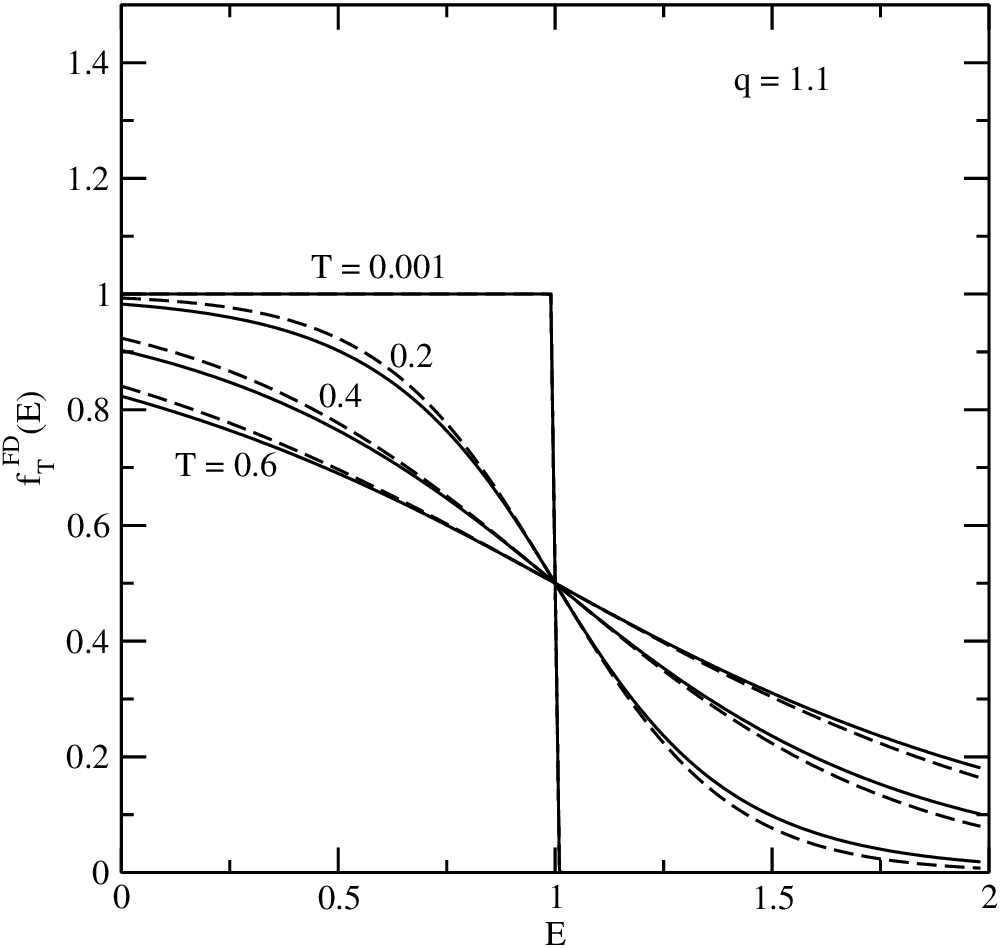}
\caption{Comparison between the Fermi-Dirac (dashed line) and 
the Tsallis-FD (solid line) distributions 
as  function of the energy $E$, keeping the Tsallis parameter $q$ fixed at 1.1,
 for 
various values of the temperature $T$. The chemical potential is kept equal to one in all curves, the units
are arbitrary.}
\label{compare-tsallis-fd}
\end{center}
\end{figure}

\begin{figure}
\begin{center}
\includegraphics[width=0.45\textwidth,height=6cm]{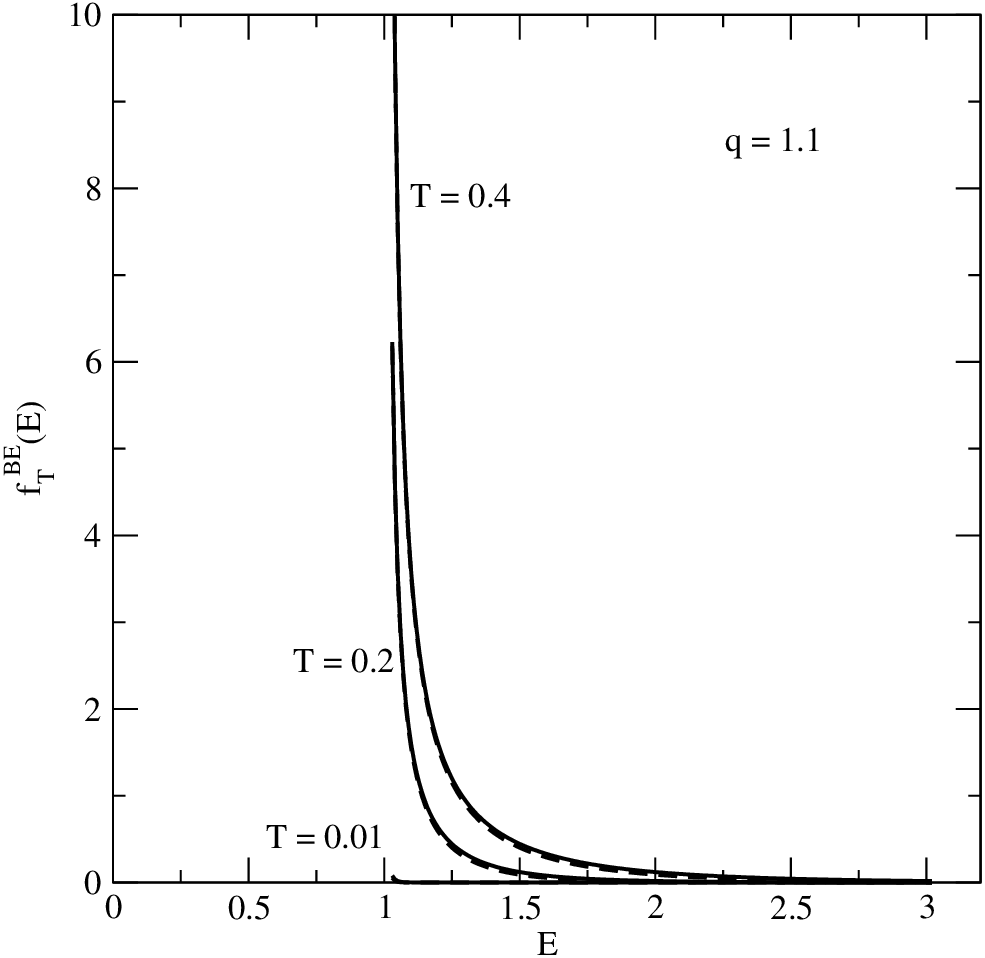}
\caption{Comparison between the Bose-Einstein (dashed line) 
and Tsallis-BE (solid line)  distributions 
as a function of the energy $E$, keeping the Tsallis parameter $q$ fixed at 1.1, for various values of the temperature $T$. 
The chemical potential is kept equal to one in all curves, the units
are arbitrary.}
\label{compare-tsallis-be}
\end{center}
\end{figure}

\begin{figure}
\begin{center}
\includegraphics[width=0.45\textwidth,height=6cm]{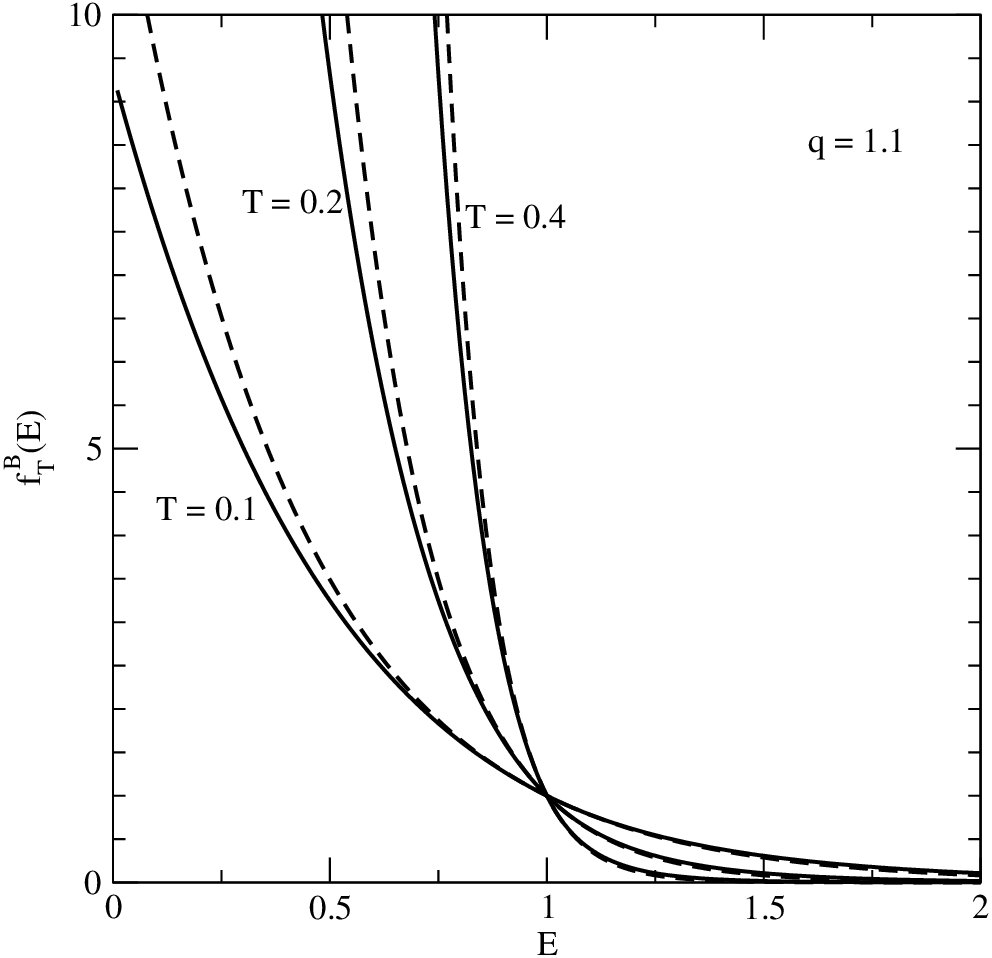}
\caption{Comparison between the Boltzmann (dashed line)
 and Tsallis-B (solid line) distributions 
as a function of the energy $E$, keeping the Tsallis parameter $q$ fixed at 1.1,
 for various values of the temperature $T$. 
The chemical potential is kept equal to one in all curves, the units
are arbitrary.}
\label{compare-tsallis-b}
\end{center}
\end{figure}

The Tsallis form of  the Fermi-Dirac distribution proposed 
in~\cite{Conroy:2010wt,turkey1,Pennini1995309,Teweldeberhan:2005wq,Conroy:2008uy} uses 
\begin{equation}
f^{FD}_T(E) \equiv 
\frac{1}{\exp_q\left(\frac{E-\mu}{T}\right)+1}  .
\label{tsallis-fd}
\end{equation}
where the function $\exp_q(x)$ is defined as
\begin{equation}
\exp_q(x) \equiv \left\{
\begin{array}{l l}
\left[1+(q-1)x\right]^{1/(q-1)}&~~\mathrm{if}~~~x > 0 \\
\left[1+(1-q)x\right]^{1/(1-q)}&~~\mathrm{if}~~~x \leq 0 \\
\end{array} \right.
\label{tsallis-fd1}
\end{equation}
and, in the limit where $q \rightarrow 1$ reduces to the standard exponential:
$$
\lim_{q\rightarrow 1}\exp_q(x)\rightarrow \exp(x).
$$
The form given in Eq.~\eqref{tsallis-fd} will be referred to as the Tsallis-FD distribution. 
A comparison between the standard Fermi-Dirac and Tsallis-FD distributions as a function of the 
energy $E$ is shown in Fig.~\ref{compare-tsallis-fd} for various values of the temperature $T$. 
The Tsallis parameter $q$ is kept fixed at $q = 1.1$.
The Bose-Einstein version will be referred to as the Tsallis-BE distribution~\cite{Chen200265} 
\begin{equation}
f^{BE}_T(E) \equiv 
\frac{1}{\exp_q\left(\frac{E-\mu}{T}\right)-1}  .
\label{tsallis-be}
\end{equation}
A comparison between the standard Bose-Einstein and Tsallis-BE distributions as a function of the energy $E$ is 
shown in Fig.~\ref{compare-tsallis-be} for various values of the 
temperature $T$. The Tsallis parameter $q$ is again
kept fixed at $q = 1.1$.
The classical limit  will be referred to as Tsallis-B  distribution (the B stands for 
the fact that it reduces to the Boltzmann distribution in the limit where $q\rightarrow 1$)
and is given by~\cite{Tsallis:1987eu,Tsallis:1998ab}
\begin{equation}
f^{B}_T(E) \equiv 
\exp_q\left(-\frac{E-\mu}{T}\right).
\label{tsallis-b}
\end{equation}
or, using standard notation,
\begin{equation}
f_T^B(E) = \left[ 1 + (q-1) \frac{E-\mu}{T}\right]^{-\frac{1}{q-1}} .
\label{tsallis}
\end{equation}
Again, a comparison between the standard Boltzmann and Tsallis-B distributions as a function of the energy $E$ 
is shown in Fig.~\ref{compare-tsallis-b}. for various values of the temperature $T$. As before the Tsallis 
parameter $q$ is kept fixed at $q = 1.1$.
All forms of the   Tsallis distribution introduce a new parameter $q$. In practice this parameter is always close to 1, 
e.g. in the results obtained by the ALICE and CMS collaborations typical values for the parameter $q$ can be obtained 
from fits to the transverse momentum distribution for identified charged particles ~\cite{Aamodt:2011zj} and are 
in the range 1.1 to 1.2 (see below). The value of $q$ should thus be considered as never being far from 1, 
deviating from it by  20\% at most. An analysis of the composition of final state particles leads to a 
similar result~\cite{Cleymans:2008mt} for the parameter $q$.

In the limit where $q\rightarrow 1$ all distributions coincide with the standard statistical distributions:
\begin{eqnarray}
\lim_{q\rightarrow 1} f_T^B(E)    &=& f^B(E), \\
\lim_{q\rightarrow 1} f_T^{FD}(E) &=& f^{FD}(E), \\
\lim_{q\rightarrow 1} f_T^{BE}(E) &=& f^{BE}(E) .
\end{eqnarray}
A derivation of the Tsallis distribution, based on the Boltzmann equation, has been given in Ref.~\cite{Biro:2005uv,Barnafoldi}.

The Tsallis-B distribution is always larger than the Boltzmann one if $q>1$. Taking into account the large $p_T$ results for 
particle production we will only consider this case here. As a consequence, in order to keep the particle yields the same, 
the Tsallis distribution always leads to smaller values of the freeze-out temperature for the same set of 
particle yields~\cite{Cleymans:2008mt}.
\subsection{Derivation for Quantum Statistics}
The standard form of the entropy for fermions in statistical mechanics is given in the large volume limit by:
\begin{eqnarray}\label{genfermi}
S^{FD}=-gV&&\displaystyle\int \frac{d^3p}{(2\pi)^3}
\left[ f^{FD}\ln f^{FD}\right.\nonumber\\
&&+\left.(1-f^{FD}) \ln(1-f^{FD}) \right],
\end{eqnarray}
For simplicity Eq.~\eqref{genfermi} refers to 
one particle species but can be easily generalized to many. In the limit where momenta are quantized this is given by:
\begin{equation}\label{genfermi-small}
S^{FD}=-g\displaystyle\sum_{i}\left[ f_i\ln f_i+(1-f_i) \ln(1-f_i) \right],
\end{equation}
For convenience we will work with the discrete form in the rest of this section. The large volume limit can be recovered 
with the standard replacement:
\begin{equation}
\displaystyle\sum_i \rightarrow V\displaystyle\int \frac{d^3p}{(2\pi)^3}
\end{equation}
The generalization, using the Tsallis prescription, leads to  
\cite{turkey1,Pennini1995309,Teweldeberhan:2005wq}
\begin{equation}\label{genfermi-tsallis}
S^{FD}_T=-g\displaystyle\sum_{i}\left[ f_i^{q}\ln_{q} f_i+(1-f_i)^{q} \ln_{q}(1-f_i) \right],
\end{equation}
where use has been made of the function 
\begin{equation}
\ln_q (x)\equiv \frac{x^{1-q}-1}{1-q} , \label{suba} 
\end{equation}
often referred to as q-logarithm. 
The classical limit of this form is given by~\cite{biro_molnar}:
\begin{equation}\label{genboltzmann-tsallis}
S^{B}_T=-g\displaystyle\sum_{i}\left[ f_i^{q}\ln_{q} f_i -f_i) \right],
\end{equation}
The equilibrium distributions can also be derived from the R\'enyi distribution
as shown in detail in~\cite{biro_van}.
It can be easily shown that in the limit where the Tsallis parameter $q$
tends to 1 one has:
\begin{equation}
\lim_{q\rightarrow 1}\ln_q (x) = \ln (x) . 
\end{equation} 
In a similar vein,  the generalized form of the  entropy 
for bosons is given by
\begin{equation}\label{genboson-tsallis}
S^{BE}_T=- g\displaystyle\sum_i\left[ f_i^q\ln_{q} f_i-(1+f_i)^q \ln_{q}(1+f_i) \right],
\end{equation}
In the limit $q\rightarrow 1$ Eqs.~\eqref{genfermi-tsallis} and 
\eqref{genboson-tsallis} reduce to the standard Fermi-Dirac and 
Bose-Einstein distributions.
Further, as we shall presently explain, the formulation 
of a variational principle in terms of  the above equations allows 
to prove the validity of the general relations of thermodynamics. 
One of the relevant constraints 
is given by the average number of particles,
\begin{equation}\label{n1}
\displaystyle\sum_{i} f_i^{q} =  N .
\end{equation}

Likewise, the energy of the system gives a constraint,
\begin{equation}\label{n2}
\displaystyle\sum_{i} f_i^{q}E_{i} = E .
\end{equation}
It is necessary to have the power $q$ on the left-hand side as no 
thermodynamic consistency would be achieved without it.
The maximization of the entropy measure 
under the constraints 
Eqs.~\eqref{n1} and \eqref{n2} leads
to the variational equation:
\begin{equation}\label{n3}
\frac{\delta}{\delta f_i}
\left[ S^{FD}_T+\alpha (N -\displaystyle\sum_i f_i^q) +\beta(E -\displaystyle\sum_i f_i^qE_i)\right] =0,
\end{equation}
where $\alpha$ and $\beta$ are Lagrange multipliers 
associated, respectively, with the total number of particles and 
the total energy. Differentiating each 
expression in Eq.~\eqref{n3} separately gives the following results
\begin{equation}\label{n4}
\frac{\delta}{\delta f_i}\left(  S^{FD}_T\right) 
=\frac{q}{q-1}\left[\left( \frac{1-f_i}{f_i}\right)^{q-1}-1 \right] f_i^{q-1},
\end{equation}
\begin{equation}\label{n5}
\frac{\delta}{\delta f_i}\left( N - \displaystyle\sum_if_i^q\right) 
= -qf_i^{q-1},
\end{equation}
and
\begin{equation}\label{n6}
\frac{\delta}{\delta f_i}\left(E - \displaystyle\sum_if_i^qE_i\right)
 =-qE_if_i^{q-1}.
\end{equation}
By substituting Eqs.~\eqref{n4}, \eqref{n5} and 
\eqref{n6} into Eq.~\eqref{n3}, we obtain
\begin{equation}\label{n7}
qf_i^{q-1} \left\lbrace \frac{1}{q-1}\left[  -1+\left(\frac{1-f_i}{f_i} \right)^{q-1} \right]
-\beta E_{i}-\alpha \right\rbrace=0.
\end{equation}
Which can be rewritten as
\begin{equation}\label{n8}
\frac{1}{q-1}\left[  -1+\left(\frac{1-f_i}{f_i} \right)^{q-1} \right]=\beta E_{i}+\alpha,
\end{equation}
and, by rearranging Eq.~\eqref{n8}, we get
\begin{displaymath}
\frac{1-f_i}{f_i}=\left[ 1+(q-1)(\beta E_{i}+\alpha)\right]^{\frac{1}{q-1}},
\end{displaymath}
which gives the Tsallis-FD form referred to earlier
in this paper
as~\cite{turkey1,Pennini1995309,Teweldeberhan:2005wq}
\begin{eqnarray}
f_i&=&\frac{1}{\left[ 1+(q-1)(\beta E_{i}+\alpha)\right]^{\frac{1}{q-1}}+1},\nonumber \\
     &=&\frac{1}{\left[\exp_{q}(\alpha +\beta E_{i} )\right] +1}.
\end{eqnarray}
Using a similar approach one can also determine the Tsallis-BE distribution by 
starting from the extremum of the entropy subject to the same two conditions:
\begin{equation}\label{n10}
\frac{\delta}{\delta f_i}\left[ S^{BE}_T+\alpha(N -\displaystyle\sum_{i} 
f_i^{q}) +\beta(E -\displaystyle\sum_{i} f_i^{q}E_{i})\right] =0,
\end{equation}
which leads to
\begin{eqnarray}
f_i&=&\frac{1}{\left[ 1+(q-1)(\beta E_{i}+\alpha)\right]^{\frac{1}{q-1}}-1},\nonumber \\
     &=&\frac{1}{\left[\exp_{q}((E_{i} -\mu)/T)\right]-1}  .\label{q}
\end{eqnarray}
where the usual identifications $\alpha = -\mu/T$  and $\beta = 1/T$ have been made. 
\subsection{\label{sec:Consistency}Thermodynamic Consistency}
%
%
The first and second laws of thermodynamics lead to the following two 
differential relations~\cite{deGroot:1980aa}
\begin{eqnarray}
 d\epsilon = Tds + \mu dn,\label{a5}\\
dP = sdT + nd\mu.\label{a51}
\end{eqnarray}
where $\epsilon = E/V$, $s = S/V$ and $n = N/V$ are the energy, entropy and particle
densities respectively.
Thermodynamic consistency 
requires that the following relations be satisfied
\begin{eqnarray}\label{a6}
 T &=& \left.\frac{\partial \epsilon}{\partial s}\right|_{n},\label{eq:a61}\\
 \mu &=&\left.\frac{\partial \epsilon}{\partial n}\right|_{s},\label{eq:a62}\\
 n &=& \left.\frac{\partial P}{\partial \mu}\right|_{T},\label{eq:a63}\\
 s &=& \left.\frac{\partial P}{\partial T}\right|_{\mu}.\label{eq:a64}
\end{eqnarray}
The pressure, energy density and entropy density are all given by corresponding
integrals over Tsallis distributions and the derivatives have 
to reproduce the corresponding physical quantities, e.g. for Tsallis-B 
one has 
\begin{eqnarray}
n_T^B &=& g\int\frac{d^3p}{(2\pi)^3}
\left[ 1 + (q-1) \frac{E-\mu}{T}\right]^{-\frac{q}{q-1}} , \\
\epsilon_T^B &=& g\int\frac{d^3p}{(2\pi)^3}E
\left[ 1 + (q-1) \frac{E-\mu}{T}\right]^{-\frac{q}{q-1}} ,\\
P_T^B &=& g\int\frac{d^3p}{(2\pi)^3}\frac{p^2}{3E}
\left[ 1 + (q-1) \frac{E-\mu}{T}\right]^{-\frac{q}{q-1}} .
\label{tsallis-nep}
\end{eqnarray}
For consistency, these expressions have to agree with the basic 
thermodynamic relations~\eqref{eq:a61},
\eqref{eq:a62},\eqref{eq:a63}, and~\eqref{eq:a64}.
i.e. , for the above relations, it has to be shown that
\begin{equation}
n_T^B = \frac{\partial P_T^B}{\partial\mu}
\end{equation}
We prove that this is indeed the case.
We will show that the consistency conditions given above are indeed 
obeyed.
%
\begin{equation}\label{a10}
 P = \frac{-E + TS + \mu N}{V}, 
\end{equation}
and take the partial derivative with respect to $\mu$ in
order to check for thermodynamic consistency, it leads to
\begin{eqnarray}\label{a11}
\left.\frac{\partial P}{\partial \mu}\right|_T & = &
\frac{1}{V}
\left[-\frac{\partial E}{\partial \mu} +T\frac{\partial S}{\partial \mu} + N + \mu\frac{\partial N}{\partial \mu}\right],\nonumber \\  
& = &\frac{1}{V}\left[N + \displaystyle\sum_{i}-\frac{T}{q-1}\left(1 + 
(q-1)\frac{E_{i} -\mu}{T}\right)\frac{\partial f_i^{q}}{\partial \mu} \right.\nonumber \\
&& \left.+\frac{Tq(1-f_i)^{q-1}}{q-1}\frac{\partial f_i}{\partial \mu}
\right],
\end{eqnarray}
then, by explicit calculation
\begin{displaymath}
 \frac{\partial f_i^{q}}{\partial \mu} = \frac{qf_i^{q+1}}{T}\left[1+(q-1)\frac{E_{i}-\mu}{T}\right]^{-1+\frac{1}{1-q}},
\end{displaymath}
\begin{displaymath}
 \frac{\partial f_i}{\partial \mu} = \frac{f_i^{2}}{T}\left[1+(q-1)\frac{E_{i}-\mu}{T}\right]^{-1+\frac{1}{1-q}},
\end{displaymath}
and
\begin{displaymath}
\left(1-f_i\right)^{q-1} = f_i^{q-1}\left[1+\frac{(q-1)(E_{i}-\mu)}{T}\right].
\end{displaymath}
Introducing this  into Eq.~\eqref{a11},  yields
\begin{equation}\label{a12}
\left. \frac{\partial P}{\partial \mu}\right|_T = n,
\end{equation}
which proves the thermodynamic consistency \eqref{eq:a63}. 

We also calculate explicitly the relation in Eq.~\eqref{eq:a61} can be rewritten as
\begin{eqnarray}\label{a13}
\left. \frac{\partial E}{\partial S}\right|_n& = &\frac{\frac{\partial E}{\partial T}dT + \frac{\partial E}{\partial \mu}d\mu}{\frac{\partial S}{\partial T}dT
 + \frac{\partial S}{\partial \mu}d\mu},\nonumber \\
& = &\frac{\frac{\partial E}{\partial T} + \frac{\partial E}{\partial \mu}\frac{d\mu}{dT}}{\frac{\partial S}{\partial T}
 + \frac{\partial S}{\partial \mu}\frac{d\mu}{dT}},
\end{eqnarray}
since $n$ is kept fixed one has the additional constraint
\begin{displaymath}
 dn = \frac{\partial n}{\partial T}dT + \frac{\partial n}{\partial \mu}d\mu = 0,
\end{displaymath}
leading to
\begin{equation}\label{a18}
 \frac{d\mu}{dT} = -\frac{\frac{\partial n}{\partial T}}{\frac{\partial n}{\partial \mu}}.
\end{equation}
Now, we rewrite \eqref{a13} and \eqref{a18} in terms of the following 
expressions
\begin{displaymath}
 \frac{\partial E}{\partial T} = \displaystyle\sum_{i} qE_{i}f_i^{q-1}\frac{\partial f_i}{\partial T}, 
\end{displaymath}
\begin{displaymath}
 \frac{\partial E}{\partial \mu} = \displaystyle\sum_{i} qE_{i}f_i^{q-1}\frac{\partial f_i}{\partial \mu}, 
\end{displaymath}
\begin{displaymath}
 \frac{\partial S}{\partial T} = \displaystyle\sum_{i} q\left[\frac{-f_i^{q-1}+(1-f_i)^{q-1}}{q-1}\right]\frac{\partial f_i}{\partial T}, 
\end{displaymath}
\begin{displaymath}
 \frac{\partial S}{\partial \mu} = \displaystyle\sum_{i}q\left[\frac{-f_i^{q-1}+(1-f_i)^{q-1}}{q-1}\right]\frac{\partial f_i}{\partial \mu}, 
\end{displaymath}
\begin{displaymath}
 \frac{\partial n}{\partial T} = \frac{1}{V}\left[\displaystyle\sum_{i} qf_i^{q-1}\frac{\partial f_i}{\partial T}\right], 
\end{displaymath}
and
\begin{displaymath}
 \frac{\partial n}{\partial \mu} = \frac{1}{V}\left[\displaystyle\sum_{i} qf_i^{q-1}\frac{\partial f_i}{\partial \mu}\right]. 
\end{displaymath}
By introducing the above relations into Eq.~\eqref{a13}, the numerator 
 of Eq.~\eqref{a13} becomes
\begin{eqnarray}\label{a19}
\frac{\partial E}{\partial T} &+& \frac{\partial E}{\partial \mu}\frac{d\mu}{dT} 
 = \displaystyle\sum_{i} qE_{i}f_i^{q-1}\frac{\partial f_i}{\partial T}\nonumber \\
&&-\frac{\displaystyle\sum_{i,j} q^{2}E_{j}\left(f_if_j\right)^{q-1}\frac{\partial f_j}{\partial \mu}\frac{\partial f_i}{\partial T}}{\displaystyle\sum_{j} 
qf_j^{q-1}\frac{\partial f_j}{\partial \mu}},\nonumber \\ 
& =& \frac{\displaystyle\sum_{i,j} qE_{i}\left(f_if_j\right)^{q-1}C_{ij}}
{\displaystyle\sum_{j} 
f_j^{q-1}\frac{\partial f_j}{\partial \mu}}.
\end{eqnarray}
Where the abbreviation 
\begin{equation}
 C_{ij}\equiv \left(f_if_j\right)^{q-1}\left[\frac{\partial f_i}{\partial T}\frac{\partial f_j}{\partial \mu}-
 \frac{\partial f_j}{\partial T}\frac{\partial f_i}{\partial \mu}\right],
\end{equation}
has been introduced. 
One can rewrite the denominator part of Eq.~\eqref{a13}  as
\begin{eqnarray}\label{a20}
\frac{\partial S}{\partial T} + \frac{\partial S}{\partial \mu}\frac{d\mu}{dT} & = 
 \frac{\displaystyle q\sum_{i,j}\left[{-f_i^{q-1}+(1-f_i)^{q-1}}\right]f_j^{q-1}
 C_{i,j}}
 {(q-1)\displaystyle\sum_{j} 
f_j^{q-1}\frac{\partial f_j}{\partial \mu}},\nonumber \\
& = \frac{\displaystyle q\sum_{i,j}(E_{i}-\mu)\left(f_if_j\right)^{q-1}
 C_{i,j}}
 {T\displaystyle \sum_{j} 
f_j^{q-1}\frac{\partial f_j}{\partial \mu}},
\end{eqnarray}
where
\begin{displaymath}
 \frac{-f_i^{q-1}+(1-f_i)^{q-1}}{q-1} = \frac{(E_{i}-\mu)}{T}f_i^{q-1},
\end{displaymath}
hence, by substituting Eqs.~\eqref{a19} and \eqref{a20} in to Eq.~\eqref{a13}, we find
\begin{equation}\label{a21}
\left. \frac{\partial E}{\partial S}\right|_n = T\frac{\displaystyle\sum_{i,j}E_{i}C_{ij}}
{\displaystyle\sum_{i,j}(E_{i}-\mu)C_{ij}},
\end{equation}
since $\displaystyle\sum_{i,j} C_{ij} = 0$, this finally leads to the desired result
\begin{equation}
\left.\frac{\partial E}{\partial S}\right|_n = T.
\end{equation}
Hence thermodynamic consistency is satisfied.
 
It has thus been shown that the definitions of temperature and
pressure within the Tsallis formalism 
for non-extensive statistics lead to expressions which
satisfy consistency with the first and second laws of thermodynamics.
The remaining relations can be be shown to be satisfied in a similar manner.
%
\section{Transverse Momentum Distributions:  Fit Details} 
\begin{figure}
\begin{center}
\includegraphics[width=0.45\textwidth,height=6cm]{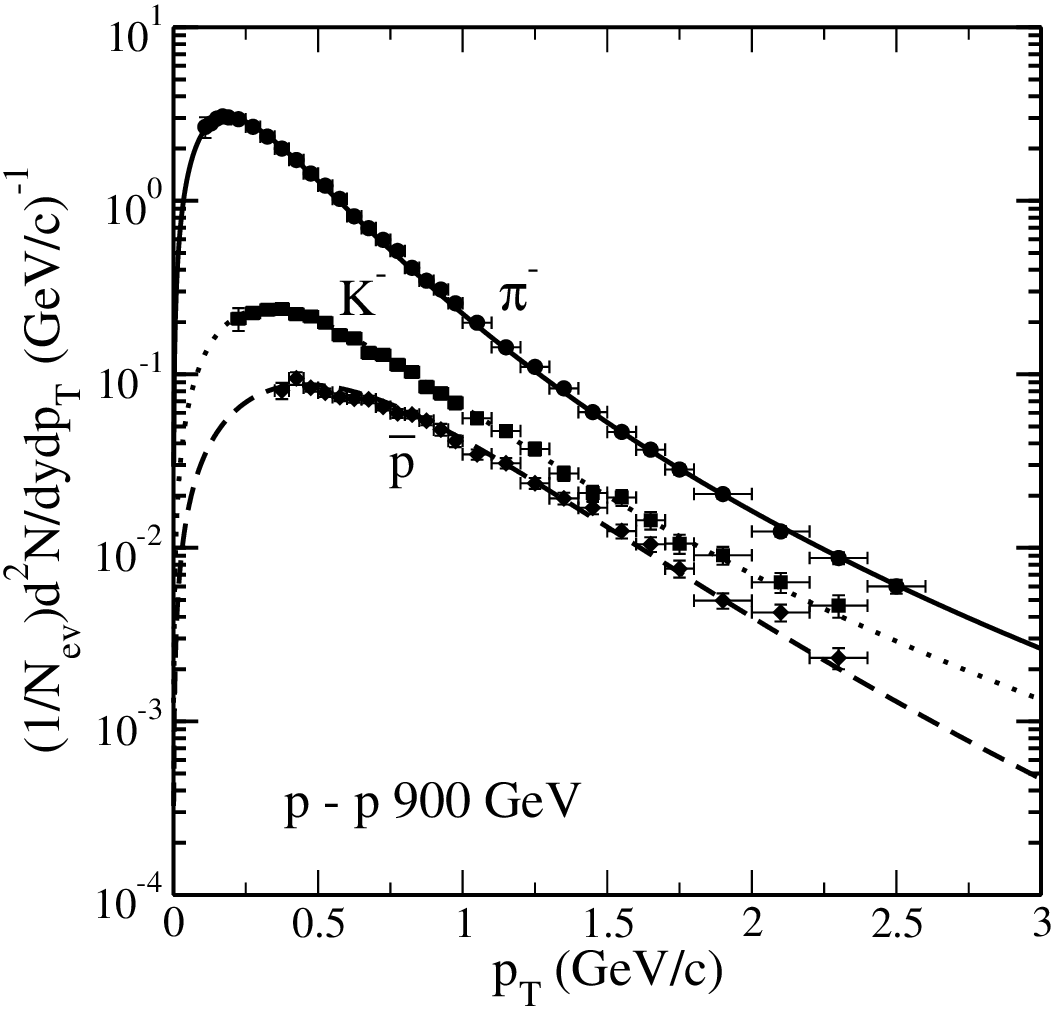}
\caption{Comparison between the 
measured transverse momentum distribution for $\pi^-$, $K^-$ and $\bar{p}$
 as measured by 
the ALICE collaboration~\cite{Aamodt:2011zj} and  the Tsallis-B distribution.
The lines are fits using the parameterization given in Eq.~\eqref{final} 
to the 0.9 TeV data with the parameters listed in Table 1. Solid line is for $\pi^-$, 
the dotted line is for $K^-$, the dashed line is for anti-protons.}
\label{fig:ALICE}
\end{center}
\end{figure}

\begin{figure}
\begin{center}
\includegraphics[width=0.45\textwidth,height=6cm]{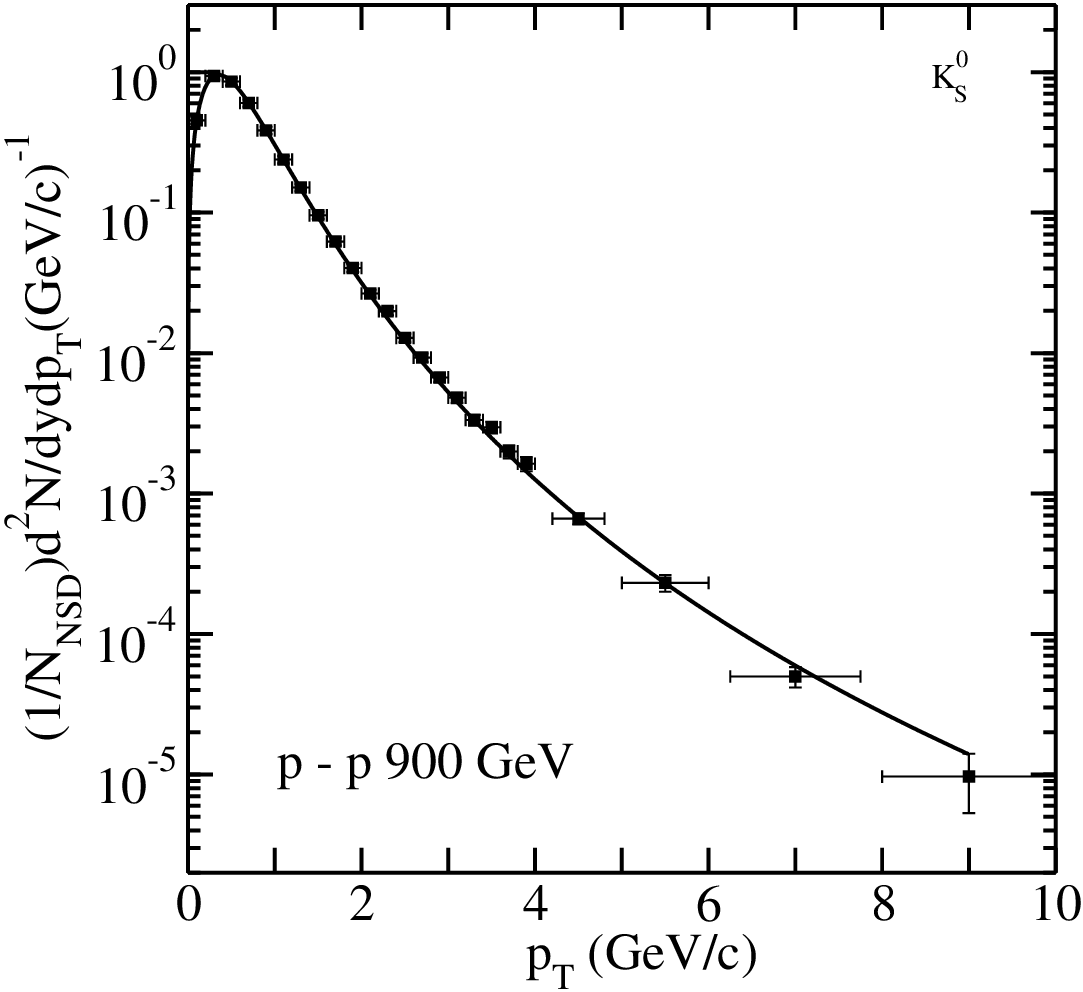}
\caption{Comparison between the 
measured transverse momentum distribution for $K^0_S$ as measured by 
the CMS collaboration~\cite{Khachatryan:2011tm} and  the Tsallis-B distribution.
The solid line is a fit using the parameterization
given in Eq.~\eqref{final} 
to the 0.9 TeV data with the parameters listed in Table 1.  
}
\label{CMS-K0S}
\end{center}
\end{figure}

\begin{figure}
\begin{center}
\includegraphics[width=0.45\textwidth,height=6cm]{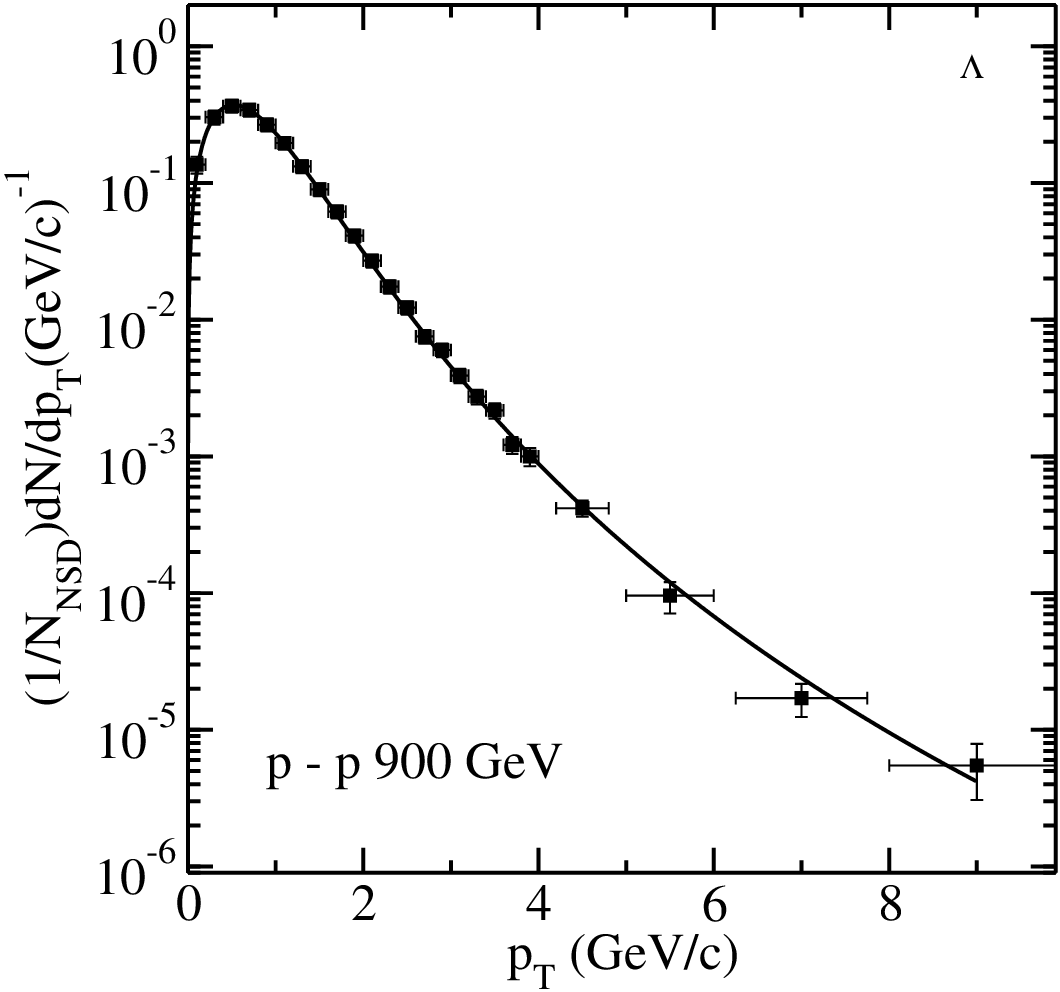}
\caption{Comparison between the 
measured transverse momentum distribution for $\Lambda$ as measured by 
the CMS collaboration~\cite{Khachatryan:2011tm} and  the Tsallis-B distribution.
The solid line is a fit using the parameterization
given in Eq.~\eqref{final} to the 0.9 TeV data with the parameters listed in Table 1.  
}
\label{fig:CMS-Lambda}
\end{center}
\end{figure}

\begin{figure}
\begin{center}
\includegraphics[width=0.45\textwidth,height=6cm]{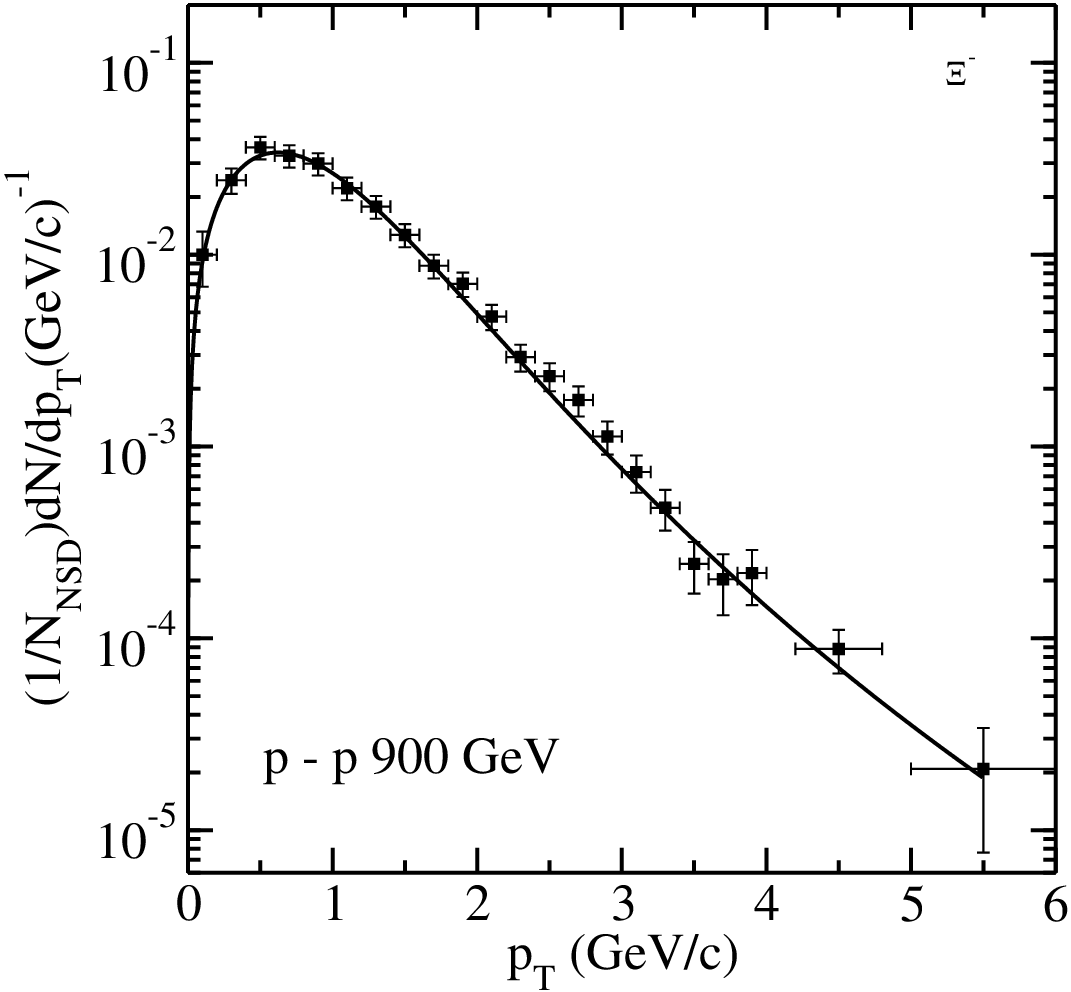}
\caption{Comparison between the 
measured transverse momentum distribution for $\Xi^-$ as measured by 
the CMS collaboration~\cite{Khachatryan:2011tm} and  the Tsallis-B distribution.
The solid line is a fit using the parameterization given in Eq.~\eqref{final}
 to the 
 0.9 TeV data with the parameters listed in Table 1.  
 }
\label{CMS-Xi}
\end{center}
\end{figure}

\begin{figure}
\begin{center}
\includegraphics[width=0.45\textwidth,height=6cm]{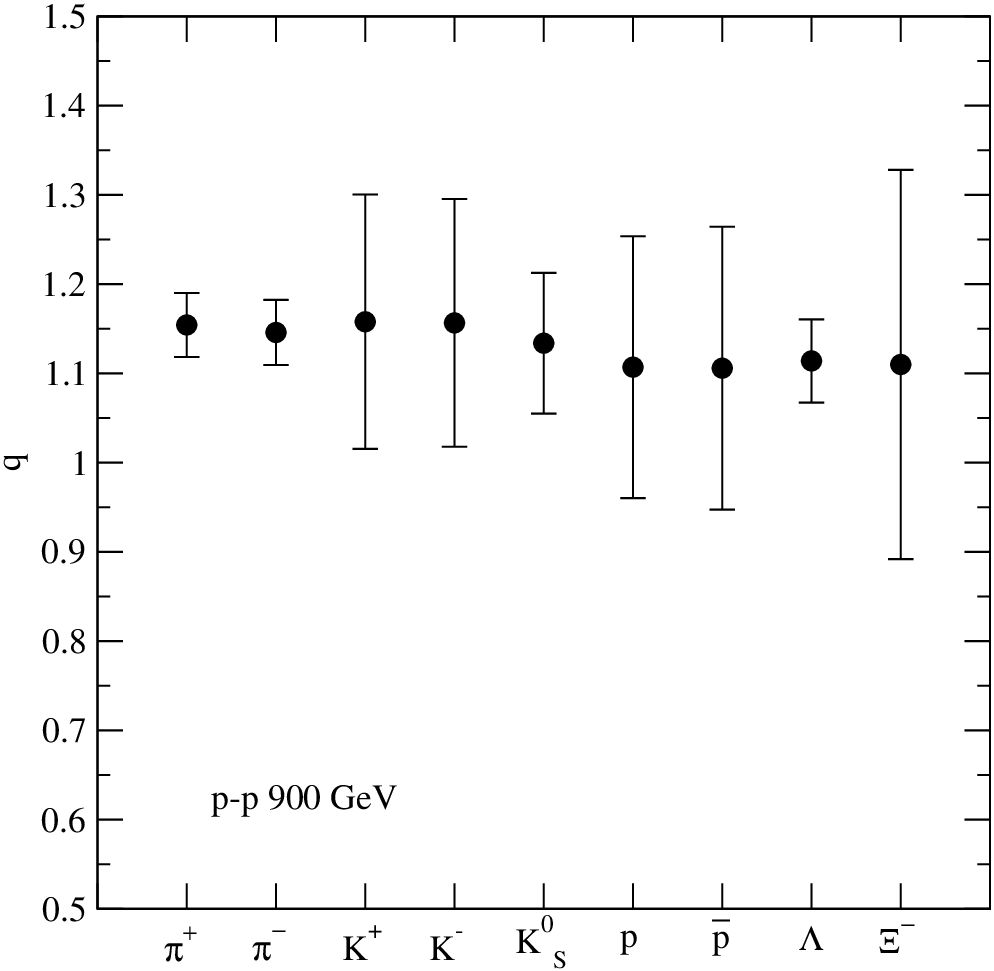}
\caption{Values of the Tsallis parameter $q$ for different  species of hadrons. }
\label{fig:q_value}
\end{center}
\end{figure}

\begin{figure}
\begin{center}
\includegraphics[width=0.45\textwidth,height=6cm]{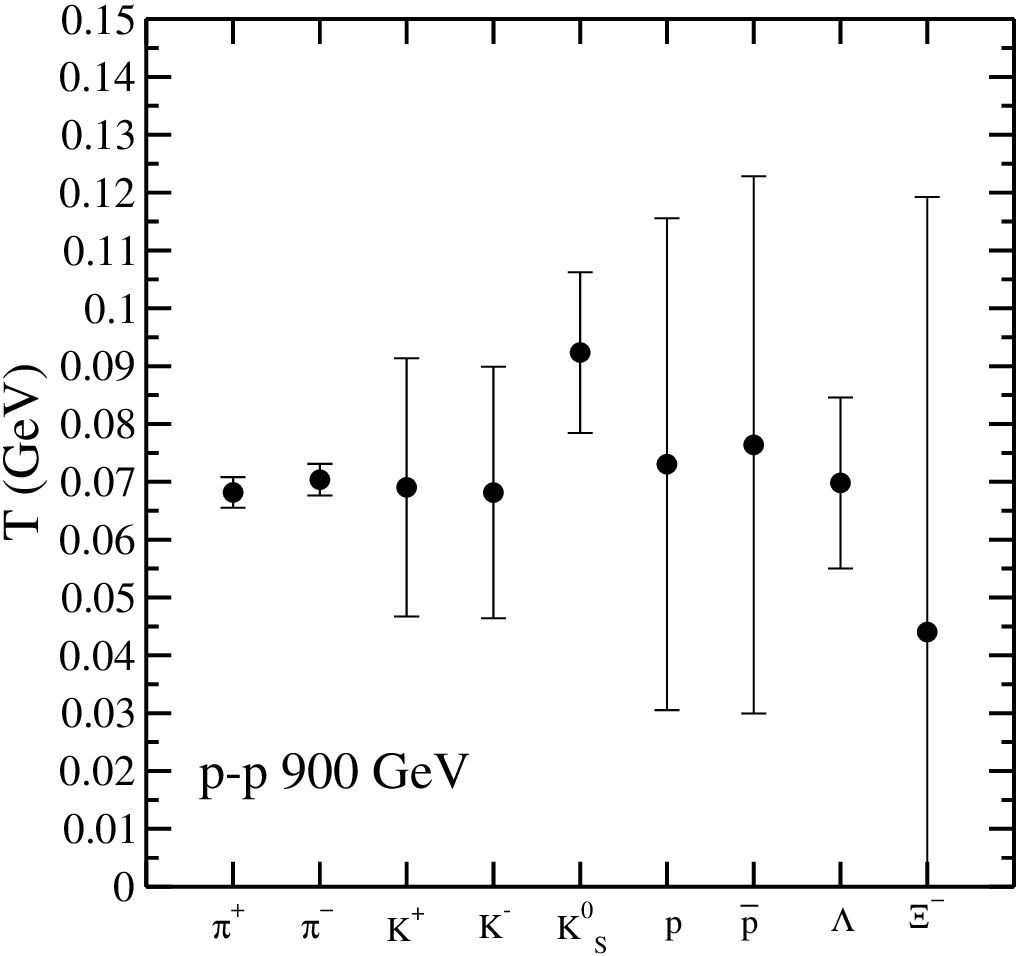}
\caption{Values of the Tsallis temperature $T$ for different species of hadrons. }
\label{fig:T_value}
\end{center}
\end{figure}

The total number of particles is given by the  integral version of Eq.~(\eqref{n1}), 
\begin{equation}
N = gV\displaystyle\int \frac{d^3p}{(2\pi)^3}
\left[1+(q-1)\frac{E-\mu}{T}\right]^{-q/(q-1)}  .
\end{equation}
The corresponding (invariant)  momentum distribution deduced from the 
equation above is given by 
\begin{equation}
E\frac{d^{3}N}{d^3p} = gVE\frac{1}{(2\pi)^3}
\left[1+(q-1)\frac{E-\mu}{T}\right]^{-q/(q-1)}.
\end{equation}
In terms of the rapidity and transverse mass variables this becomes

\begin{eqnarray}
\frac{d^{2}N}{dy\, p_Tdp_T} &=& gV\frac{m_T\cosh y}{(2\pi)^2}\nonumber\\
&&\times\left[1+(q-1)\frac{m_T\cosh y -\mu}{T}\right]^{-q/(q-1)}.
\end{eqnarray}
At mid-rapidity $y=0$ and for zero chemical potential this reduces to
the following expression
\begin{equation}
\left.\frac{d^{2}N}{dp_T~dy}\right|_{y=0} = gV\frac{p_Tm_T}{(2\pi)^2}
\left[1+(q-1)\frac{m_T}{T}\right]^{-q/(q-1)}.
\label{final}
\end{equation}
In Fig.~\ref{fig:ALICE} we show a fit to the 
transverse momentum distributions obtained in $p-p$ collisions at 900 GeV
for identified particles $\pi^-, K^-, \bar{p}$ published by the ALICE 
collaboration~\cite{Aamodt:2011zj}.
We have also checked that the $\chi^2$ values are of a similar
quality.
We have also made fits using the Tsallis-B distribution
 to experimental 
measurements published by the CMS collaboration~\cite{Khachatryan:2011tm}.
These are shown in Figs.~\ref{CMS-K0S},~\ref{fig:CMS-Lambda} and~\ref{CMS-Xi} 
and are comparable 
with those 
shown by the CMS  collaboration~\cite{Khachatryan:2011tm}. 
The resulting parameters  are collected in Table 1. The most striking feature is that the values of the 
parameter $q$ are fairly stable in the range 1.1 to 1.2 for all particles considered at 0.9 TeV. 
The temperature $T$ cannot be determined very accurately for all hadrons but they are consistent with a value around 70 MeV.\\
%
%
%
\begin{table}[ht]
\begin{center}
\begin{tabular}{|c|c|c|c|}
\hline
Particle & $q$                &$T$ (GeV)      & $\chi^2$/ndf\\
\hline 
$\pi^+$  & 1.154  $\pm$0.036 & 0.0682 $\pm $0.0026 & 12.01/30\\ 
$\pi^-$  & 1.146  $\pm$0.036 & 0.0704 $\pm$ 0.0027 &13.28/30\\
$K^+$    & 1.158  $\pm$0.142 & 0.0690 $\pm $0.0223  &16.25/24\\
$K^-$    & 1.157  $\pm$0.139 & 0.0681 $\pm$ 0.0217  &7.06/24\\
$K^0_S$  & 1.134  $\pm$0.079 & 0.0923 $\pm $0.0139  &14.41/21\\
$p$      & 1.107  $\pm$0.147 & 0.0730 $\pm$ 0.0425  & 14.77/21\\
$\bar{p}$& 1.106  $\pm$0.158 & 0.0764 $\pm $0.0464  &13.18/21\\
$\Lambda$& 1.114  $\pm$0.047 & 0.0698 $\pm$ 0.0148  & 8.45/21\\
$\Xi^-$  & 1.110  $\pm$0.218 & 0.0440 $\pm$ 0.0752  & 10.09/21\\
\hline  
\end{tabular}
\caption{Fitted values of the $T$ and $q$ parameters for different species
of hadrons measured by the ALICE~\cite{Aamodt:2011zj} and 
CMS collaborations~\cite{Khachatryan:2011tm}, together with
the corresponding $\chi^2$ values,  
using the Tsallis-B form for the momentum distribution. 
}
\end{center}
\label{table1}
\end{table}
For clarity we show these results also in Fig.~\ref{fig:q_value} for 
the values of 
the parameter $q$ and in Fig.~\ref{fig:T_value} for the values
 of the Tsallis parameter $T$.
The striking feature is that the values of $q$ are consistently between 1.1 and 1.2
for all species of hadrons at 0.9.
%
%
%
The values obtained for the temperature are clearly below values for the 
thermal freeze-out temperature that have been reported elsewhere in the
literature. This is unavoidable when using the Tsallis distribution: 
for the same value of $T$, the Tsallis distribution is always higher than the 
Boltzmann distribution, hence, to reproduce the same transverse momentum,
one has to use a lower temperature for the Tsallis distribution
than for the Boltzmann one.

\begin{figure}
\begin{center}
\includegraphics[width=0.45\textwidth,height=6cm]{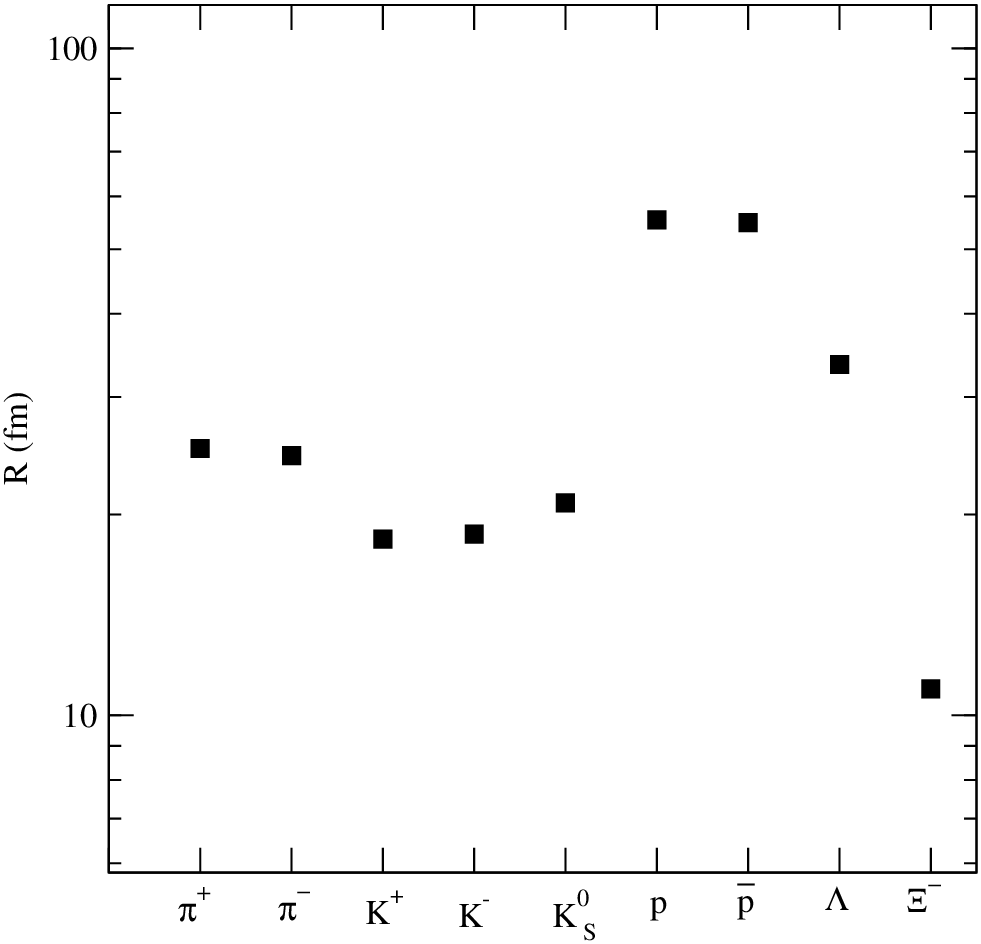}
\caption{Values of the radius $R$ deduced from the volume $V$ 
appearing in \eqref{eq:Tsallis-B} for different particle species in
$p-p$ collisions at 900 GeV.}. 
\label{fig:r_value}
\end{center}
\end{figure}
the ALICE~\cite{ALICE_all}, ATLAS~\cite{aad} and CMS~\cite{CMS_all} 
collaborations  have published data on the transverse momentum distribution
of charged particles. These extend to much higher
values of the transverse momentum~\cite{aad} and would provide
 an important test for
distinguishing formula~\eqref{eq:Tsallis-B} from~\eqref{ALICE-CMS}. 
 Since this does not involve identified particles, 
it makes use of  a summation over several hadrons, e.g. 
pions, kaons and protons, hence the analysis is a bit more involved and will be considered in a 
separate publication.\\
For completeness we also show the value of the volume $V$ appearing 
in Eq.~\eqref{eq:Tsallis-B}. The resulting radius $R$ is shown 
in Fig.~\ref{fig:r_value}. If all hadrons originate from the same system
and if there were no extra contributions coming from heavier resonances
decaying into hadrons, then this volume should be the same for all hadrons. 
This is clearly not the case in the present analysis, in fact the radius is 
surprisingly larger which can only be interpreted by a very large time
between chemical and thermal freeze-out. This clearly needs further investigation.
\section{Discussion and Conclusions}
In this paper we have presented a detailed derivation  of the quantum form
of the Tsallis distribution and proven 
the thermodynamic consistency of the resulting distribution.
It was emphasized that an additional power of $q$ is needed to achieve
consistency with the laws of thermodynamics~\cite{Conroy:2010wt}.
The resulting distribution was compared with 
recent measurements from the ALICE~\cite{Aamodt:2011zj} 
and CMS collaborations~\cite{Khachatryan:2011tm}
and good agreement was obtained. The resulting parameter $q$ which is
a measure for the deviation from a standard Boltzmann distribution was found
to be in the range 1.1-1.2.  The resulting values of the temperature are
also consistent within the  errors and lead to a value of 
around 70 MeV.  
The analysis presented here  cannot be considered complete as several
elements are still missing. Most important is the contribution
of heavier resonance which contribute 
to the final number of pions and kaons. Their effect on heavier baryons
like the $\Xi$ is not as large as for pions but nevertheless it is a factor
that has to be taken into account. This could  change  the
conclusions presented here.

\end{document}